\documentstyle[psfig]{aa}  
\begin{document}

   \thesaurus{09     
              (03.20.9 VLT;  
	       08.08.1;  
	       08.16.5;  
               09.08.1 NGC3603;  
               11.19.3)} 
   \title{Low-mass stars in the massive H\,II region
NGC\,3603\thanks{Based on observations obtained at the European Southern
Observatory, Paranal (ESO Proposal ID 63.I-0015)}}

   \subtitle{Deep NIR imaging with ANTU/ISAAC}

   \author{B. Brandl
        \inst{1}
        \and
        W. Brandner
	\inst{2}
	\and
	F. Eisenhauer
	\inst{3}
	\and
	A.F.J. Moffat
	\inst{4}
	\and
	F. Palla
	\inst{5}
	\and
	H. Zinnecker
	\inst{6}
          }

   \offprints{B. Brandl}

   \institute{Cornell University, Department of Astronomy,
              222 Space Sciences Building, Ithaca, NY 14853, USA\\
              email: brandl@astrosun.tn.cornell.edu
         \and
         University of Hawaii, Institute for Astronomy, 2680 Woodlawn Dr., 
		Honolulu, HI 96822, USA
         \and
         Max-Planck-Institut f\"ur Extraterrestrische Physik, 
		Giessenbachstra\ss e, D-85740 Garching, Germany
         \and
         D\'epartement de physique, Universit\'e de Montr\'eal, C.P. 6128, 
		Succ. Centre-ville, Montr\'eal, QC, H3C 3J7, Canada, and 
		Observatoire du mont M\'egantic
         \and
         Osservatorio Astrofisico di Arcetri, Largo E.Fermi 5, 
		I-50125 Firenze, Italy
         \and
         Astrophysikalisches Institut Potsdam, An der Sternwarte 16,
		D-14482 Potsdam, Germany
         }

   \date{Received September 9, 1999; accepted October 11, 1999}

   \maketitle

   \begin{abstract}

We have observed NGC~3603, the most massive visible H\,II region known in 
the Galaxy, with ANTU(VLT1)/ISAAC in the near-infrared (NIR) $J_{\rm s}$, 
$H$, and $K_{\rm s}$-bands.  Our observations are the most sensitive 
observations made
to date of this dense starburst region, allowing us to investigate with
unprecedented quality its low-mass stellar population.  Our mass limit to 
stars detected in all three bands is $0.1 M_\odot$ for a pre-main sequence 
star of age 0.7~Myr.  The overall age of the pre-main sequence stars in the 
core region of NGC 3603 has been derived from isochrone fitting in the
colour-magnitude diagram, leading to $0.3 - 1.0$~Myr.  The NIR luminosity
functions show that the cluster is populated in low-mass stars at least down
to $0.1 M_\odot$.  Our observations clearly show that sub-solar mass stars 
do form in massive starbursts.

   \end{abstract}

%
%
\section{Introduction}
NGC~3603 is located in the Carina spiral arm (RA = 11$^h$, 
DEC = -61$^\circ$) at a distance of $\sim 6-7$~kpc
(De Pree et al. 1999, and references therein).  It is the only
massive, Galactic H\,II region whose ionizing central cluster can
be studied at optical wavelengths due to only moderate (mainly
foreground) extinction of $A_{\rm V}\approx 4.5^{\rm m}$ (Eisenhauer et al. 
1998).  The OB stars ($\ge 10 M_\odot$)
contribute more than $2000 M_\odot$ to the cluster mass.  In
comparison to the Orion Trapezium system, NGC~3603 with its more than 50
O and WR stars (Moffat, Drissen \& Shara 1994)
producing a Lyman continuum flux of $10^{51} \mbox{s}^{-1}$ 
(Kennicutt 1984; Drissen et al. 1995) has about 100 times the ionizing power
of the Trapezium cluster.
With a bolometric luminosity $L_{\rm bol} > 10^7 L_\odot$, NGC~3603 has 
about 10\% of the luminosity of 30~Doradus and looks in many respects very 
similar to its stellar core R136 (Brandl et al. 1996).  In fact
it has been called a Galactic clone of R136 without the
massive surrounding cluster halo (Moffat, Drissen \& Shara 1994).
In many ways NGC~3603 and R136 can be regarded as representative 
building blocks of more distant and luminous starburst galaxies
(Brandl, Brandner \& Zinnecker 1999, and references therein).

\begin{figure*}
\vbox{\psfig{figure=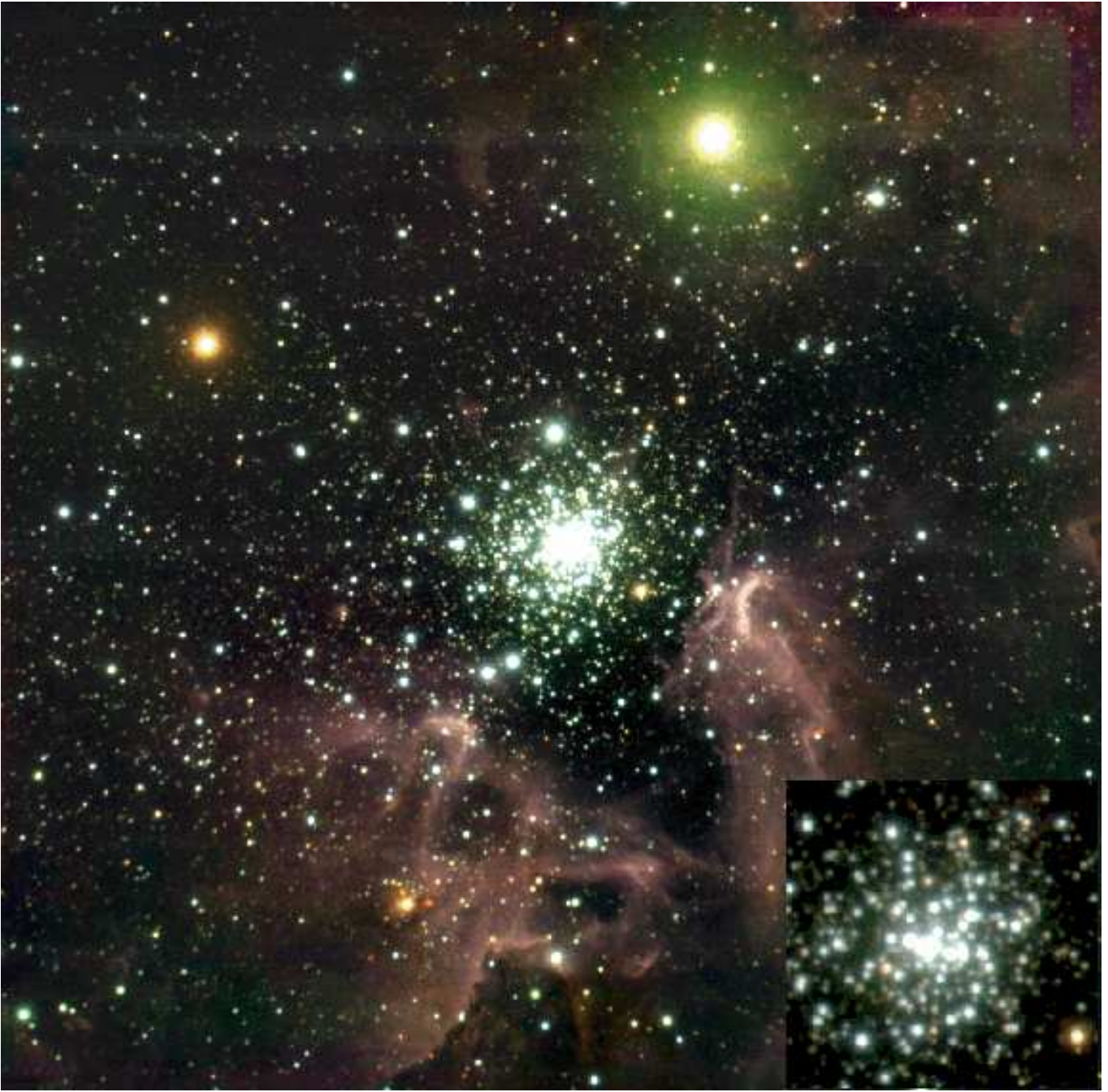,width=18.0cm}\vspace{-0cm}}
\hfill\parbox[b]{18cm}{\caption[]{Three colour image of NGC~3603 composed
		from $J_{\rm s}$ (blue), $H$ (green), and $K_{\rm s}$-band 
		(red) images.
		Intensities are scaled in logarithmic units; FOV is 
		$3.\!'4 \times 3.\!'4$ ($6.2 \times 6.2$ parsec$^2$). 
		North is up, East to the left.  The insert to the lower 
		right is a blow up of the central parsec$^2$.}
\label{vltimage}}%
\end{figure*}
%

Although NGC~3603 has been the target of many recent ground and space 
based studies (Eisenhauer et al. 1998; Brandner et al.  2000) little 
is known about its content of sub-solar mass stars.
Our general aim is to investigate low-mass star formation in the violent
environments of starburst regions (Brandl, Brandner \& Zinnecker 1999).
Several fundamental questions arise in this context: 
Does the slope of the IMF vary on small scales?  
Do low-mass stars in a starburst event form together with the most massive 
stars or do they form at different times or on different timescales?  
And finally, one might even ask if low-mass stars form at all in such 
environments.

\section{Observations and Data Reduction}
NGC~3603 was observed in the $J_{\rm s} = 1.16 - 1.32\mu$m, 
$H = 1.50 - 1.80\mu$m, and $K_{\rm s} = 2.03 - 2.30\mu$m broadband filters 
using the NIR camera ISAAC mounted on ANTU, ESO's first VLT.  
The observations were made during the 4 nights of 
April 4 -- 6 and 9, 1999, in service mode when the optical seeing 
was equal or better than $0.\!''4$ on Paranal.  Such seeing was essential for
accurate photometry in the crowded cluster and increased our sensitivity
to the faintest stars.  The majority of our data were taken under
photometric conditions.

\begin{figure*}
\vbox{\psfig{figure=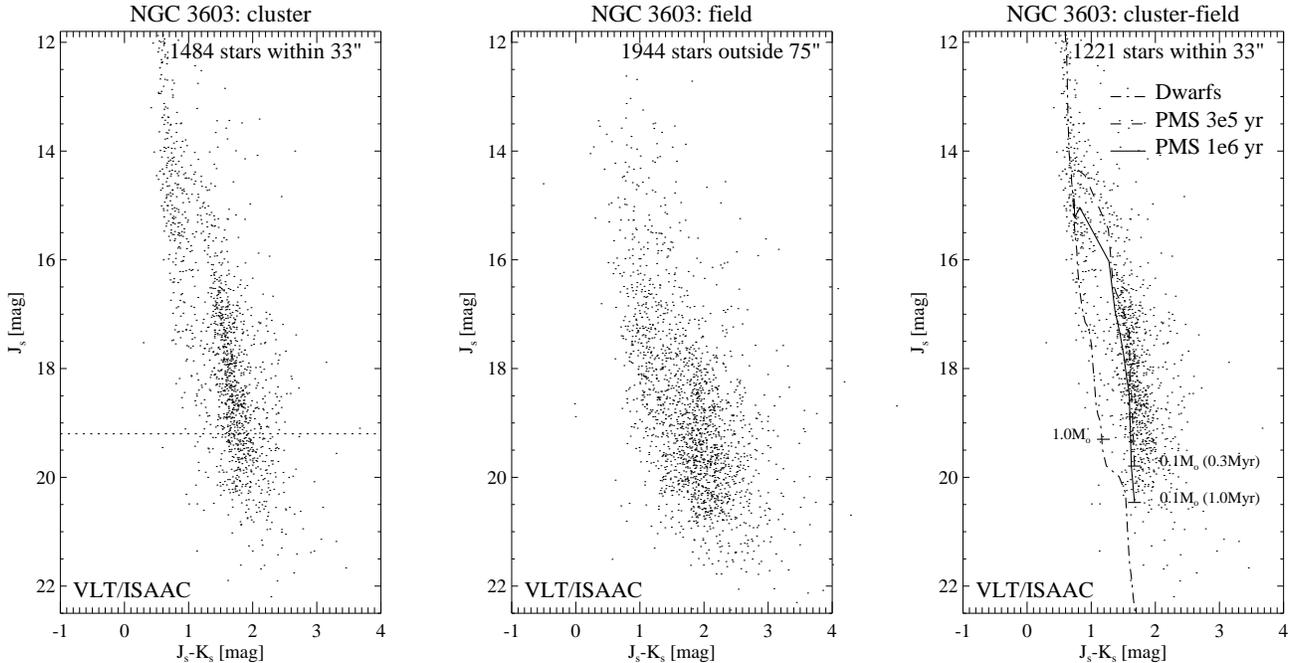,width=18.0cm,angle=90}\vspace{-0.2cm}}
\hfill\parbox[b]{18cm}{\caption[]{{\bf a-c.} 
		$J_{\rm s}$ versus $J_{\rm s} - K_{\rm s}$ colour-magnitude 
		diagrams of NGC~3603. {\bf a} contains all stars
		detected in all three wavebands within the central 
		$r \le 33''$ (1\,pc) [$1'^\Box$], {\bf b} shows the 
		field stars at $r \ge 75''$ (2.25\,pc) [$6.3'^\Box$] 
		around the cluster, and {\bf c} shows the cluster population
		within $r \le 33''$ with the field stars 
                statistically subtracted.
		The dashed horizontal line indicates the detection limit of 
		the previous most sensitive NIR study by Eisenhauer et al. 
		(1998). 
		{\bf c} also shows the theoretical isochrones of pre-main
		sequence stars of different ages from Palla \& Stahler 
		(1999) and the main sequence for dwarfs. 
		For comparison we've plotted some corresponding stellar
		masses next to the isochrones.}  
\label{cmds}}%
\end{figure*}

Our observing strategy was to use the shortest possible frame times
(1.77s) to keep the number of saturated stars to a minimum.
However, due to the system's excellent sensitivity, about two dozen of
the brightest stars ended up being saturated.  Nevertheless, this does
not impose a problem to our study of the low mass stars.
Thirty-four short exposures were co-added to an effective one-minute
exposure, the minimum time per pointing required to stabilize the 
telescope's active optics control system.  Between the 1 minute pointings 
we moved the telescope by up to $20''$ offsets in a random pattern. This
approach has several advantages:
\begin{itemize}
\item{Enlargement of the observed field of view (FOV) with maximum 
	signal-to-noise (S/N) in the cluster center.}
\item{Reduction of residual images and other
	array artefacts, using the median filtering technique.}
\item{Derivation of the ``sky'' from the target exposures using the median
	filtering technique. No additional time for ``blank'' sky 
	frames outside the cluster was required.}
\end{itemize}  
The sky frames have been computed using between 15 and 37 subsequent 
exposures per waveband and night, and careful eye-inspection showed that all
sources have been efficiently removed using our modified median filtering
technique which returns the lower 1/3 instead of the mean (1/2) value.
We subtracted the sky-background and flat-fielded each exposure
using the twilight flat-fields provided by ESO. The relative position
offsets were derived from cross-correlating the images; the exposures
were co-aligned on a $0.5\times 0.5$-pixel subgrid for better spatial
resolution, and then added together using the median filtering technique.  
The resulting images are $3.\!'4 \times 3.\!'4$ in size with pixels 
of $0.\!''074$.  The
effective exposure times of the final broadband images in the central
$2.\!'5 \times 2.\!'5$ is 37, 45, and 48 minutes in $J_{\rm s}$, $H$, and
$K_{\rm s}$, respectively. Fig.~\ref{vltimage} shows the impressive 3-colour
composite image.  
The brightest star in the FOV ($80''$ northeast of the core) is the red 
supergiant IRS~4 (Frogel, Persson, \& Aaronson 1977).
The ring nebula and the bipolar outflows around the blue supergiant 
Sher~25 (Brandner et al. 1997) about $18''$ north of the core are 
clearly detected in the K~band.  
The image also shows the three proplyd-like objects that have
been recently discovered by Brandner et al. (2000); these proplyds
are similar to those seen in Orion but about 20-30 times more extended.
About $1'$ south of the central cluster, we detect the brightest members of
the deeply embedded proto cluster IRS~9.

In order to derive the photometric fluxes of the stars we used the IRAF
implementation of DAOPHOT (Stetson 1987). We first ran DAOFIND to detect
the individual sources, leading to $\sim$20,000 peaks in each waveband.
Many of these may be noise or peaks in the nebular background and
appear only in one waveband.  In order to reject spurious sources, we
required that sources be detected independently in all three wavebands,
and that the maximal deviation of the source position centroid between
different wavebands be less than 0.075$''$.  The resulting source list
contains 6967 objects in the entire FOV.  

\begin{figure*}
\vbox{\psfig{figure=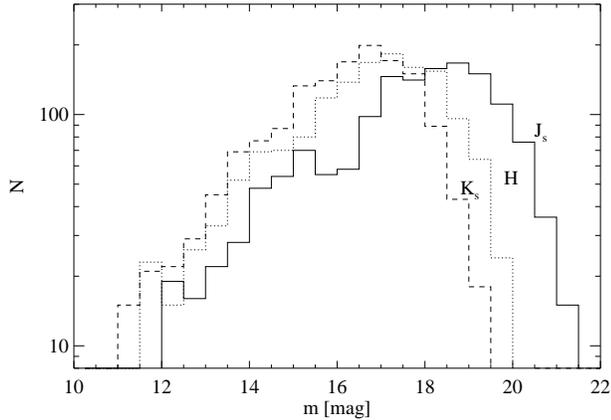,width=8.8cm,angle=90}\vspace{-0.0cm}}
\vspace{-0.4cm}
\parbox[b]{8.8cm}{\caption[]{The observed luminosity functions of stars
		detected at all 3 wavebands in the 
		central stellar cluster with $r \le 33''$ (1\,pc). The
		magnitudes are apparent $J_{\rm s}$ (solid), $H$ (dotted), 
		and $K_{\rm s}$ (dashed) magnitudes.  The number counts
		have not been corrected for incompleteness.}
\label{lumif}}%
\end{figure*}

We then flux-calibrated the images using the faint NIR standard stars from 
the lists by Hunt et al. (1998) and Persson et al. (1998).  Because of the 
stringent requirements on the seeing the PSF did not noticeably change 
during our observations and the systematic photometric errors are dominated 
by uncertainties in the aperture offsets.  (A detailed error analysis will 
be part of a subsequent paper).  Comparing our photometric fluxes of
numerous sources with the fluxes derived by Eisenhauer et al.\ (1998)
yields a systematic offset of $0.1^{\rm m}$ in $J_{\rm s}$ and 
$0.05^{\rm m}$ in $K_{\rm s}$.

\section{Results and Interpretation}
Fig.~\ref{cmds}{\bf a-c.} shows the resulting colour-magnitude diagrams 
(CMD). The left plot contains all stars detected in all 3 wavebands within 
$r \le 33''$ (1~pc).  Since NGC~3603 is located in the Galactic Plane we
expect a significant contamination from field stars.  To reduce this 
contribution we followed a statistical approach by subtracting the
average number of field stars found in the regions around the cluster at 
$r \ge 75''$ (central plot) per magnitude and per colour bin (0.5 mag each).
The accuracy of our statistical subtraction is mainly limited by three 
factors: 
first, we cannot rule out that low-mass pre-main sequence stars are also 
present in the outskirts of the cluster. Second, because of crowding on one 
hand and dithering which leads to shorter effective integration times 
outside the central $2.\!'5 \times 2.\!'5$ on the other hand, the 
completeness limit varies across the FOV.  
Third, local nebulosities may hide background
field stars.  However, none of these potential errors affects our
conclusions drawn from the CMD.

The resulting net CMD for cluster stars within $r \le 33''$ of NGC~3603 is 
shown at the right in Fig.~\ref{cmds}.  We overlayed the theoretical 
isochrones of pre-main sequence stars from Palla \& Stahler (1999) down to 
$0.1 M_\odot$. We assumed a
distance modulus of $(m-M)_{\rm o} = 13.9$ based on the distance of 
6~kpc (De Pree et al. 1999) and an average foreground extinction of 
$A_{\rm V} = 4.5^{\rm m}$ following the reddening law by Rieke \& Lebofski 
(1985).
Applying the isochrones to measured magnitudes could be misleading since
the theoretical calculations include only stellar photospheres while
the stars still may be surrounded by dust envelopes and accretion disks.
This would lead to excess emission in the NIR and make the stars appear
younger than they actually are.  Typical excess emission of classical T
Tauri stars in the Taurus--Auriga complex have been determined as
$\Delta H = 0.2^{\rm m}$, and $\Delta K = 0.5^{\rm m}$ (Meyer, Calvet \& 
Hillenbrand 1997).  
The upper part of the cluster-minus-field CMD clearly shows a main
sequence with a marked knee indicating the transition to pre-main
sequence stars.  The turn-on occurs at 
$J_{\rm s}\approx 15.5^{\rm m}$ ($m\approx 2.9 M_\odot$).  Below the
turn-on the main-sequence basically disappears.  We note that the width
of the pre-main sequence in the right part of Fig.~\ref{cmds} does not
significantly broaden toward fainter magnitudes, indicating that our
photometry is not limited by photometric errors.  The scatter may in fact
be real and due to varying foreground extinction, infrared excess and 
evolutionary stage.  In that case the left rim of the distribution would 
be representative of the ``true'' colour of the most evolved stars 
while the horizontal scatter would be primarily caused by accretion disks 
of different inclinations and ages.  
Fitting isochrones to the left rim in the CMD yields an age of only 
$0.3 - 1.0$~Myr.  Our result is in good agreement with the study by 
Eisenhauer et al. (1998) but extends the investigated mass range by about 
one order of magnitude toward smaller masses.  
Because of $\approx 10$ magnitudes range in luminosities in the crowded 
core region our sensitivity limits of $J_{\rm s} \approx 21^{\rm m}$, 
$H \approx 20^{\rm m}$ and $K_{\rm s} \approx 19^{\rm m}$ don't appear to be 
exceedingly faint (and are about 3 magnitudes above ISAAC's detection limit 
for isolated sources).
However, only VLT/ISAAC's high angular resolution, PSF stability, and 
overall sensitivity enabled us to study the 
sub-solar stellar population in a starburst region on a star by star basis. 

Fig.~\ref{lumif} shows the luminosity functions for stars detected in all 
3 wavebands.  For the purpose of this letter we have not attempted to 
correct our number counts for incompleteness, i.e., an increasingly 
significant percentage of stars will be undetected toward fainter 
magnitudes. Thus we have no means to determine whether the apparent 
turnover at $K_{\rm s}\approx 16.5^{\rm m}$ is real or an observational 
artifact, but we do note that the mass spectrum is well populated down to 
$K_{\rm s}\sim 19^{\rm m}$, corresponding to a $0.1 M_\odot$ pre-main 
sequence star of 0.7~Myr (Zinnecker, McCaughrean \& Wilking 1993).  
A detailed analysis of the low-mass IMF will be the subject of a subsequent 
paper.

\begin{acknowledgements}
      We would like to thank the ESO staff for their excellent work.
\end{acknowledgements}

\end{document}